\documentclass[amsmath,amssymb,superscriptaddress,twocolumn]{revtex4-2}

\usepackage{graphicx}
\usepackage{dcolumn}

\def\be{\begin{equation}}
\def\ee{\end{equation}}
\def\beq{\begin{eqnarray}}
\def\eeq{\end{eqnarray}}

\usepackage{bm}
\usepackage{graphicx}
\usepackage{dcolumn}
\usepackage{amsmath}
\usepackage[latin1]{inputenc}
\usepackage{graphicx, psfrag}
\usepackage{amssymb}
\usepackage[colorlinks=true, citecolor=blue, urlcolor = blue, linkcolor= red, bookmarks=true]{hyperref}
\usepackage{float}
\usepackage{amsmath}
\usepackage{amsfonts}
\usepackage{dcolumn}
\usepackage{hyperref}
\usepackage{subfigure}
\usepackage{pgfplots}
\usepackage{epstopdf}
\usepackage{booktabs}

\begin{document}


\title{Wormholes and energy conditions in $f(R,T)$ gravity}



\author{Ayan Banerjee} \email[]{ayanbanerjeemath@gmail.com}
\affiliation{Atrophysics Research Centre, School of Mathematics, Statistics and Computer Science, University of KwaZulu--Natal, Private Bag X54001, Durban 4000, South Africa}

\author{Takol Tangphati} 
\email[]{takoltang@gmail.com}
\affiliation{School of Science, Walailak University, Thasala, \\Nakhon Si Thammarat, 80160, Thailand}

\author{Anirudh Pradhan}
\email{pradhan.anirudh@gmail.com}
\affiliation{Centre for Cosmology, Astrophysics and Space Science, GLA University, Mathura-281 406, Uttar Pradesh, India}


\date{\today}

\begin{abstract}
We explore the existence of wormholes in the context of $f(R,T)$ gravity. The $f(R,T)$ theory is a curvature-matter coupled modified gravity that depends on an arbitrary function of the Ricci scalar $R$ and the trace of the stress-energy tensor $T$. In this work, we adopt two different choices for the matter Lagrangian density  ($\mathcal{L}_m= \mathcal{P}$ and $\mathcal{L}_m= p_r$) and investigate the impact of each one on wormhole structure. By adequately specifying the redshift function and the shape function, we found a variety of exact wormhole solutions in the theory. Our finding indicates that, for both classes of wormholes the energy density is always positive throughout the spacetime, while
the radial pressure is negative. This means exotic matter is necessary for the existence of wormholes  in $f(R,T)$ gravity.

\end{abstract}


\maketitle


\section{Introduction}\label{intro:Sec}

Wormhole acts as a hypothetical tunnel-like structure that connects two parallel universes or distant parts of the same universe. The original idea of a wormhole solution was discovered by Albert Einstein and Nathan Rosen \cite{Einstein:1953tkd} in 1935. That's why wormholes are sometimes called ``Einstein-Rosen bridges",
while the term wormhole was first introduced by Wheeler in 1957 \cite{Fuller1957}. A heuristic point of view  toward the traversable wormhole was first pointed out by Morris and Thorne \cite{Morris:1988cz}. After the seminal work, physicists have considered the wormhole solution as a possibility tool for interstellar travel connecting two asymptotically flat spacetimes. More interestingly,  authors in \cite{Morris1988} have
discussed an exciting possibility that a wormhole can be converted into a time machine.  The interested reader may see Refs. \cite{Visser:1995,Lobo:2007zb} for more information. 

In a search for theoretical models of traversable wormholes one needs an exotic matter. In the context of GR, such form of matter violates the null energy condition (NEC) \cite{Morris:1988cz,Visser:1995} at least in a neighborhood of the wormhole throat. Such wormholes may also violate the pointwise and averaged energy conditions \cite{Lobo:2007zb}. In addition to this, authors have devoted their studies to minimize the use of exotic matter. Among them ``volume integral quantifier" is the most promising way to quantify the total amount of exotic matter \cite{Visser:2003yf,Kar:2004hc}. 
This idea was further improved by Nandi \emph{et al} \cite {Nandi:2004ku} who quantifies the exact amount of exotic matter present in a given spacetime. In particular, Poisson and Visser \cite{Poisson:1995sv} developed a particular class of wormholes called thin-shell wormholes. In this setting wormholes are 
constructed by cutting and pasting two manifolds to obtain a new manifold. In those wormholes the exotic
matter is concentrated on the throat, and therefore less exotic matter will require for wormhole construction \cite{Visser:1989kh,Visser:1989kg,Lobo:2003xd,Dias:2010uh}. Thus, in the last few years many efforts have been given to find wormhole solutions in supported of exotic matter sources 
\cite{Lobo:2005us,Sushkov:2005kj,Carvente:2019gkd,Sharif:2014opa}, and see also Refs. \cite{Jamil:2010ziq,Zaslavskii:2005fs,Bronnikov:2006pt,Gonzalez:2009cy,Cataldo:2008ku} for further development.

However, it is always challenging in wormhole physics that could in principle be created by an ordinary matter (i.e., satisfy the energy conditions). In this respect, several strategies have been adopted to alleviate the problem. In particular,  higher-dimensional cosmological wormholes  \cite{Zangeneh:2014noa} and wormhole solutions in the alternative theories of gravity
\cite{Mehdizadeh:2015jra,Mazharimousavi:2016npo,Bronnikov:2002rn} have attracted  much attention. Indeed, in \cite{Pavlovic:2014gba,Lobo:2009ip} it was shown that in $f(R)$ gravity wormholes can be theoretically constructed without resorting to exotic matter. Otherwise, one can search for wormholes in different modified gravity theories, as for example $4D$ Einstein-Gauss-Bonnet gravity \cite{Jusufi:2020yus,Godani:2022jwz},
third-order Lovelock gravity \cite{KordZangeneh:2015dks,Mehdizadeh:2016nna}, Horndeski theories \cite{Bakopoulos:2021liw},
hybrid metric-Palatini gravity \cite{Rosa:2021yym,KordZangeneh:2020ixt}, $f(Q)$ gravity \cite{Banerjee:2021mqk,Parsaei:2022wnu,Hassan:2022hcb}, Extended theories of gravity \cite{DeFalco:2021klh,DeFalco:2021ksd} and other theories. 

Among several viable candidates for modified theories of gravity, we shall study the wormhole solutions in spirit of $f(R, T)$ gravity theory. In other words, $f(R, T)$ theory allows a coupling between matter and geometry \cite{Harko:2011kv} that depends on 
an arbitrary function of the Ricci scalar $R$ and of the trace of the stress-energy tensor $T$. Such gravity theory was proposed to explain the late time acceleration of the Universe  without invoking any additional matter fields, such as bulk viscosity or scalar fields. Such an extensions of general relativity (GR) includes new type of contributions of the material stress to the right-hand side of the Einstein equations. In this context, cosmological solutions have been studied in Refs. \cite{Houndjo:2011fb,jamil2011re,Jamil:2012pf,Baffou:2013dpa,Singh:2013bpa,Sharif:2014ioa,Baffou:2017pao,Mishra:2017sdq}). The authors of Ref. \cite{Noureen:2014xua} have studied spherically symmetric collapsing stars surrounding in locally anisotropic environment in $f(R, T)$ gravity. In this regard, solutions to relativistic compact stars have been widely investigated Refs.~\cite{Moraes2016,Das:2016mxq,Deb:2017rhd,Biswas:2018inc,Lobato2020,Pretel2021,Pappas:2022gtt}. The author in Ref. \cite{Ordines:2019sjq} have 
investigated changes in Earth's atmospheric models coming from the $f(R, T)$ gravity. 

Beside that wormhole solutions were also investigated in the same context \cite{Moraes:2017mir,Elizalde:2018frj,Moraes:2019pao}.
Interestingly authors in \cite{Zubair:2019uul,Rosa:2022osy,Banerjee:2020uyi} have shown that wormholes solutions could be possible without exotic matter for the entire spacetime in $f(R,T)$ gravity. Also, a charged wormholes in $f(R,T)$ gravity has been proposed recently in \cite{Moraes:2017rrv}. Considering the existence of a conformal Killing symmetry, wormhole solution has also been studied in \cite{Banerjee:2019wjj}. Here we consider two different choices of matter Lagrangian density (which is not unique) and derive their corresponding  field equations.  The main purpose of this paper is to present static and spherically symmetric wormhole solutions using those different field equations.
Moreover, we are interested to see the differences in their structures and properties.

This paper is organized in the following manner: after a brief introduction in Section \ref{intro:Sec}, we
consider the action and write out the gravitational field
equations for $f(R,T)$ theory in Section \ref{secII}. We present the field equations for analytic wormhole solutions  assuming two different choices for the matter Lagrangian density in Section \ref{sec3}. We present the structure equation for standard energy conditions in \ref{sec4}. In \ref{sec5}, we will then construct analytic wormhole solutions for specific choice of redshift and shape function.  Finally we give our conclusions in section \ref{sec6}.

\section{Basic formalism of $f(R,T)$ gravity}
\label{secII}
Modifications to the Einstein-Hilbert action take many forms, but here we are interested to introduce an arbitrary coupling between geometry and matter. Following this direction and Ref. \cite{Harko:2011kv}, we start with the action for $f(R,T)$ gravity theory which is 
\begin{equation}\label{1}
    S = \frac{1}{16\pi}\int f(R,T)\sqrt{-g}d^4x + \int\mathcal{L}_m \sqrt{-g}d^4x ,
\end{equation}
where $g$ is the determinant of the  physical metric and $f(R,T)$ is an arbitrary function  of the  $R$ (Ricci scalar) and $T$ (trace of the energy-momentum tensor), respectively. Furthermore, $\mathcal{L}_m$ is the matter field's action.

The variation of (\ref{1}) with respect to the metric $g_{\mu\nu}$, the corresponding field equations in $f(R,T)$ gravity can be written as
\begin{align}\label{2}
    f_R(R,T) R_{\mu\nu} &- \dfrac{1}{2}f(R,T) g_{\mu\nu} + [g_{\mu\nu}\square - \nabla_\mu\nabla_\nu] f_R(R,T)  \nonumber  \\
    &= 8\pi T_{\mu\nu} -(T_{\mu\nu} + \Theta_{\mu\nu})f_T(R,T) ,
\end{align}
where $R_{\mu\nu}$ is the Ricci tensor and $T_{\mu\nu}$ the energy-momentum tensor, respectively. Also,  we have defined the partial derivatives of $f$ as  $f_R \equiv \partial f/\partial R$, $f_T \equiv \partial f/\partial T$, $\square \equiv \nabla_\mu\nabla^\mu$ is the d'Alembertian operator with $\nabla_\mu$ stands for the covariant derivative. In this relation the auxiliary tensor $\Theta_{\mu\nu}$ is defined as  
\begin{align}\label{3}
    \Theta_{\mu\nu} &\equiv g^{\alpha\beta}\frac{\delta T_{\alpha\beta}}{\delta g^{\mu\nu}}  \nonumber  \\
    &=  -2T_{\mu\nu} + g_{\mu\nu}\mathcal{L}_m - 2g^{\alpha\beta} \frac{\partial^2\mathcal{L}_m}{\partial g^{\mu\nu} \partial g^{\alpha\beta}} .
\end{align}

As of $f(R)$ gravity \cite{Sotiriou:2008rp,DeFelice:2010aj,Nojiri:2010wj,Nojiri:2017ncd}, the Ricci scalar is treated as a redundant degree of freedom
in $f(R,T)$ theories and taking the trace of the field equations (\ref{2}) leads to 
\begin{align}\label{4}
    3\square f_R(R,T) &+ Rf_R(R,T) - 2f(R,T)  \nonumber  \\
    &= 8\pi T - (T+\Theta)f_T(R,T) ,
\end{align}
where we have denoted $\Theta= \Theta_\mu^{\ \mu}$. 

In addition to this, the divergence of the stress-energy tensor $T_{\mu\nu}$ (\ref{2}) can then be written as \cite{BarrientosO:2014mys}
\begin{align}\label{5}
    \nabla^\mu T_{\mu\nu} =&\ \frac{f_T(R,T)}{8\pi - f_T(R,T)}\bigg[ (T_{\mu\nu} + \Theta_{\mu\nu})\nabla^\mu \ln f_T(R,T)   \nonumber  \\
    & + \nabla^\mu\Theta_{\mu\nu} - \frac{1}{2}g_{\mu\nu}\nabla^\mu T \bigg] .
\end{align}

Note that $f(R,T)$ theory is a curvature-matter coupling theory, and therefore it leads to  a non-conservation of the energy-momentum tensor. For the purpose of wormhole solution, one has to specify the functional form of $f(R,T)$ gravity. 
Now, in this work we consider the simplest form of $f(R,T)$ which is $f(R,T)= R+ 2\beta T$ \cite{Harko:2011kv}.  As a consequence, Eqs.~(\ref{2}), (\ref{4}) and (\ref{5}) can be written as follows
\begin{align}
    G_{\mu\nu} &= 8\pi T_{\mu\nu} + \beta Tg_{\mu\nu} - 2\beta(T_{\mu\nu} + \Theta_{\mu\nu}) ,   \label{6}   \\ 
    R &= -8\pi T - 2\beta(T- \Theta) ,   \label{7}    \\
    \nabla^\mu T_{\mu\nu} &= \frac{2\beta}{8\pi - 2\beta} \left[ \nabla^\mu \Theta_{\mu\nu} - \frac{1}{2}g_{\mu\nu}\nabla^\mu T \right] ,   \label{8}
\end{align}
where $G_{\mu\nu}$ is the Einstein tensor.



\section{The wormhole geometry and the field equations} \label{sec3}

We will focus on the simplest case for a wormhole solution which can be described by a static spherically symmetric line element 
\begin{equation}\label{metric}
ds^2= -e^{2\Phi(r)}dt^2+\frac{dr^2}{1-\frac{b(r)}{r}}+r^2(d\theta^2+\sin^2\theta d\phi^2),
\end{equation}
which is the so-called Morris and Thorne metric \cite{Morris:1988cz}. Since, the two functions
$\Phi(r)$ and $b(r)$ are defined as the redshift and the shape functions with certain restrictions.  The radial coordinate $r$  decreases from $+\infty$ to a minimum value $r_0$, and then  increases from $r_0$ to $-\infty$. The minimum surface area is called the throat of the wormhole $b(r_0)= r_0$. Moreover, the flaring-out condition of the shape of the wormhole is another fundamental ingredient,  given by the condition $\frac{b(r)-rb^{\prime}(r)}{b^2(r)}>0$ \cite{Morris:1988cz}. It can
also be found that the form function satisfies the condition $b^{\prime}(r_0) < 1$. The shape function $b(r)$ should obey the following condition $1-b(r)/r > 0$ for the region out of the throat.  In addition, the restriction of finiteness on $\Phi(r)$ is also imposed to ensure the absence of horizons and singularities throughout the spacetime. 

We further assume that the energy-momentum tensor of an anisotropic fluid is of the form 
\begin{equation}\label{eq7}
T_{\mu\nu}=(\rho+p_t)u_\mu u_\nu+ p_t g_{\mu\nu}-\sigma \chi_{\mu}\chi_{\nu},
\end{equation}
which will be taken into account for wormhole matter distribution. Here, $\rho = \rho(r)$ is the energy density, $p_r = p_r(r)$ is the
radial pressure and $p_t = p_t (r)$ is the transverse pressure. The $u^\mu$ is the (timelike) four-velocity and $\chi^{\mu}$ is the unit spacelike vector in the radial direction. The anisotropy factor $\sigma = p_t-p_r$.

In the case of a isotropic/anisotropic fluid, the choice of matter Lagrangian density is not unique. It could be a function of both density and pressure i.e., $\mathcal{L}_m$ = $\mathcal{L}_m (\rho, p)$ or one could choose either $\mathcal{L}_m = p$ or $\mathcal{L}_m = -\rho$ \cite{Harko:2010zi,Faraoni:2009rk,Bertolami:2008ab}. With the above taken into consideration, we focus on  $\mathcal{L}_m = p$ and $\mathcal{L}_m = -\rho$, and discuss two different cases separately.

\subsection{Modified TOV equations for $\mathcal{L}_m = \mathcal{P}$}

As discussed in the previous section, the widely used matter Lagrangian density is given by 
$\mathcal{L}_m = \mathcal{P}$, where $\mathcal{P} \equiv (p_r+ 2p_t)/3$ (see Refs.~\cite{Deb:2018sgt,Maurya:2019sfm,Biswas:2020gzd}
for more details). Under this consideration, $\Theta_{\mu\nu} = -2T_{\mu\nu} + \mathcal{P}g_{\mu\nu}$ the Eqs.~(\ref{6}), (\ref{7}) and (\ref{8}) become  
\begin{align}
    G_{\mu\nu} &= 8\pi T_{\mu\nu} + \beta Tg_{\mu\nu} + 2\beta(T_{\mu\nu} - \mathcal{P}g_{\mu\nu}) ,   \label{11}   \\ 
    R &= -8\pi T - 2\beta(3T- 4\mathcal{P}) ,   \label{12}    \\
    \nabla^\mu T_{\mu\nu} &= \frac{2\beta}{8\pi + 2\beta} \left[ \nabla^\mu \left(\mathcal{P}g_{\mu\nu} \right) - \frac{1}{2} \nabla_\nu T \right] .   \label{13}
\end{align}
Then, using Eqs. (\ref{metric}) and (\ref{eq7}), in the field equations (\ref{11}) we find the following
set of equations
  \begin{eqnarray}
      && \frac{b^{\prime}}{r^{2}} = 8\pi\rho + \beta\left[ 3\rho - p_r - \frac{2}{3}\sigma \right] ,  \label{14}  \\
      && 2\left(1-\frac{b}{r}\right)\frac{\Phi^{\prime}}{r}-\frac{b}{r^{3}} = 8\pi p_r + \beta\left[ -\rho + 3p_r+ \frac{2}{3}\sigma \right],  \label{15}  \\
      &&  \left(1-\frac{b}{r}\right)\left[\Phi^{\prime\prime}+\Phi^{\prime 2}-\frac{b^{\prime}r-b}{2r(r-b)}\Phi^{\prime}-\frac{b^{\prime}r-b}{2r^2(r-b)} +\frac{\Phi^{\prime}}{r}\right]  \nonumber\\
      && = 8\pi(p_r + \sigma) + \beta\left[ -\rho + 3p_r+ \frac{8}{3}\sigma \right] ,  \label{16}
  \end{eqnarray}
where the prime stands for differentiation with respect to $r$. 

\subsection{Modified TOV equations for $\mathcal{L}_m = p_r$}

If one considers the matter Lagrangian is $\mathcal{L}_m = p_r$, then the 
three non-zero components are
\begin{eqnarray}
    && \frac{b^{\prime}}{r^{2}} = 8\pi\rho + \beta\left[ 3\rho - p_r - 2 \sigma  \right] , \label{eq17} \\
    && 2\left(1-\frac{b}{r}\right)\frac{\Phi^{\prime}}{r}-\frac{b}{r^{3}} = 8\pi p_r + \beta\left[ -\rho + 3p_r+ 2\sigma  \right], \label{eq18} \\
    &&  \left(1-\frac{b}{r}\right)\left[\Phi^{\prime\prime}+\Phi^{\prime 2}-\frac{b^{\prime}r-b}{2r(r-b)}\Phi^{\prime}-\frac{b^{\prime}r-b}{2r^2(r-b)} +\frac{\Phi^{\prime}}{r}\right]  \nonumber\\
    && = 8\pi(p_r + \sigma) + \beta\left[ -\rho + 3p_r+ 4\sigma \right]. \label{eq19}
\end{eqnarray}

Summarizing, we see that both sets of equations for different choices of Lagrangian density have five unknown quantities, i.e., $\Phi(r)$, $b(r)$, $\rho(r)$, $p_r(r)$ and $p_t(r)$ with three independent differential equations.
Hence, it is an undetermined system of equations, and we have to apply elegant strategies to construct wormhole solutions. We will discuss further the solutions  in the next section. 
\section{Energy Conditions} \label{sec4}

As discussed in the introduction, a traversable wormhole necessarily violates the null energy condition (NEC) (in fact, it violates all the energy conditions \cite{Visser:1995}) in the context of GR. Thus,  matter which violates the standard energy conditions is called exotic matter. But an interesting feature comes out in the framework of modified gravity theories that wormhole configurations may be constructed without exotic matter. Thus, we now turn to the question of whether our wormhole solution does satisfy energy conditions or not.  For the given anisotropic matter distribution, the NEC, weak energy condition (WEC) and strong energy condition (SEC) could be  translated into the following three constraints on the matter variables:
\begin{equation}\label{eq20}
\rho+p_i \ge 0,~\text{where}~ i = r, t,
\end{equation}
while the WEC entails
\begin{equation}\label{eq20}
\rho \ge 0 ~\text{and} ~\rho+p_i\ge 0,
\end{equation}
and the SEC asserts that
\begin{equation}\label{eq20}
\rho +\sum_{i} p_i \ge 0 ~\text{and for each} ~\rho+p_i\ge 0.
\end{equation}
In the next section, we will  exploit the above arguments while constructing a wormhole solution.

\section{Specific Cases and Wormhole Solution } \label{sec5}
In constructing wormhole solutions, we analyze specific cases for different choices of the functions $\Phi(r)$ and $b(r)$.

\subsection{Solutions with $\Phi^{\prime}(r) = 0$ and $b(r)= r_0 \sqrt{\frac{r}{r_0}}$}
\begin{figure}[h]
    \centering
    \includegraphics[width = 8.5 cm]{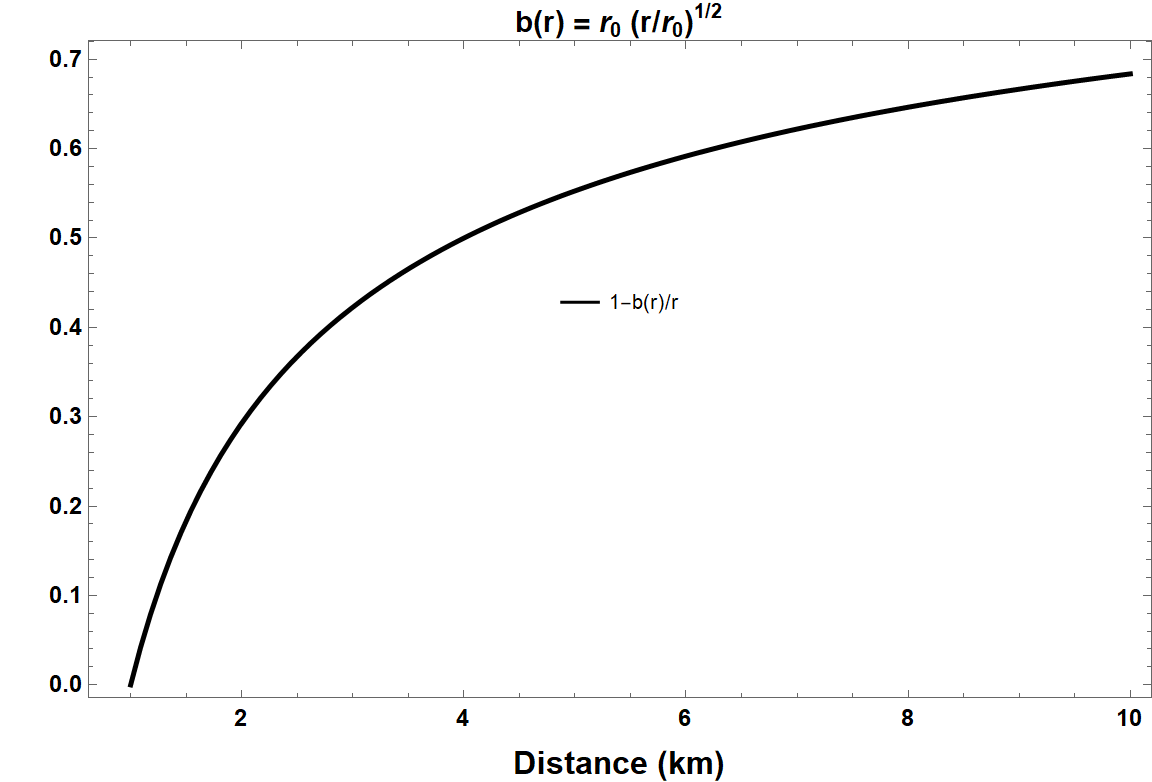}
    \caption{Profile for $1-b(r)/r$ for the specific case of $r_0 = 1$.}
    \label{figf1}
\end{figure}

Consider for simplicity a zero redshift function and a
specific choice for the shape function (see Fig. \ref{figf1}), we obtain the stress-energy tensor profile 
for two sets of equations. The specific choice for the shape function is well known and obeys all the restricted conditions as mentioned in \ref{sec3}, see Ref. \cite{Bouhmadi-Lopez:2014gza} for more.

\subsubsection{For $\mathcal{L}_m = \mathcal{P}$}
With the following Lagrangian density $\mathcal{L}_m = \mathcal{P}$ and using the Eqs. (\ref{14}-\ref{16}), we obtain
\begin{eqnarray}
    \rho(r) &=& \frac{3 \pi + \beta}{6(2 \pi + \beta)(4 \pi + \beta)} \frac{1}{r^2}\sqrt{\frac{r_0}{r}}, \\
    p_r(r) &=& - \frac{12 \pi + 5 \beta}{12(2 \pi + \beta)(4 \pi + \beta)} \frac{1}{r^2} \sqrt{\frac{r_0}{r}}, \\
    p_t(r) &=& \frac{6\pi + 5 \beta}{24(2\pi + \beta)(4 \pi + \beta)} \frac{1}{r^2} \sqrt{\frac{r_0}{r}}.
\end{eqnarray}
For the above case, the NEC along the radial and tangential direction are given by 
\begin{eqnarray}
    \rho+p_r &=& -\frac{1}{4r^2(4 \pi + \beta)}\sqrt{\frac{r_0}{r}}, \\
     \rho+p_t  &=&  \frac{3}{8r^2(4 \pi + \beta)}\sqrt{\frac{r_0}{r}},
\end{eqnarray}
respectively. For the values considered in Fig. \ref{f1}, we plot $\rho$, $\rho+p_r$, $\rho+p_t$ and $\rho+p_r+2p_t$, respectively. Observing the Fig. \ref{f1} (upper panel), we see that energy density is positive throughout the spacetime.
Furthermore, we note that $\rho+p_r <0$ and $\rho+p_t >0$ for $r_0=1$ and $\beta = -\pi$. Moreover, the situation is reversed when $\beta <-4 \pi$.
This indicates that the NEC is violated along the radial/tangential direction depending on the values of $\beta$.  

The NEC at the throat is given by $(\rho+ p_r)|_{r_0} = -\frac{1}{4r_0^2(4 \pi + \beta)}$ and 
$(\rho+p_r)|_{r_0} = \frac{3}{8r_0^2(4 \pi + \beta)}$, respectively. It seems that the NEC is violated for the normal matter threading
the throat of the wormhole.  

\vspace{0.5cm}

\subsubsection{For $\mathcal{L}_m = p_r$}

\begin{figure}[h]
    \includegraphics[width = 8.5 cm]{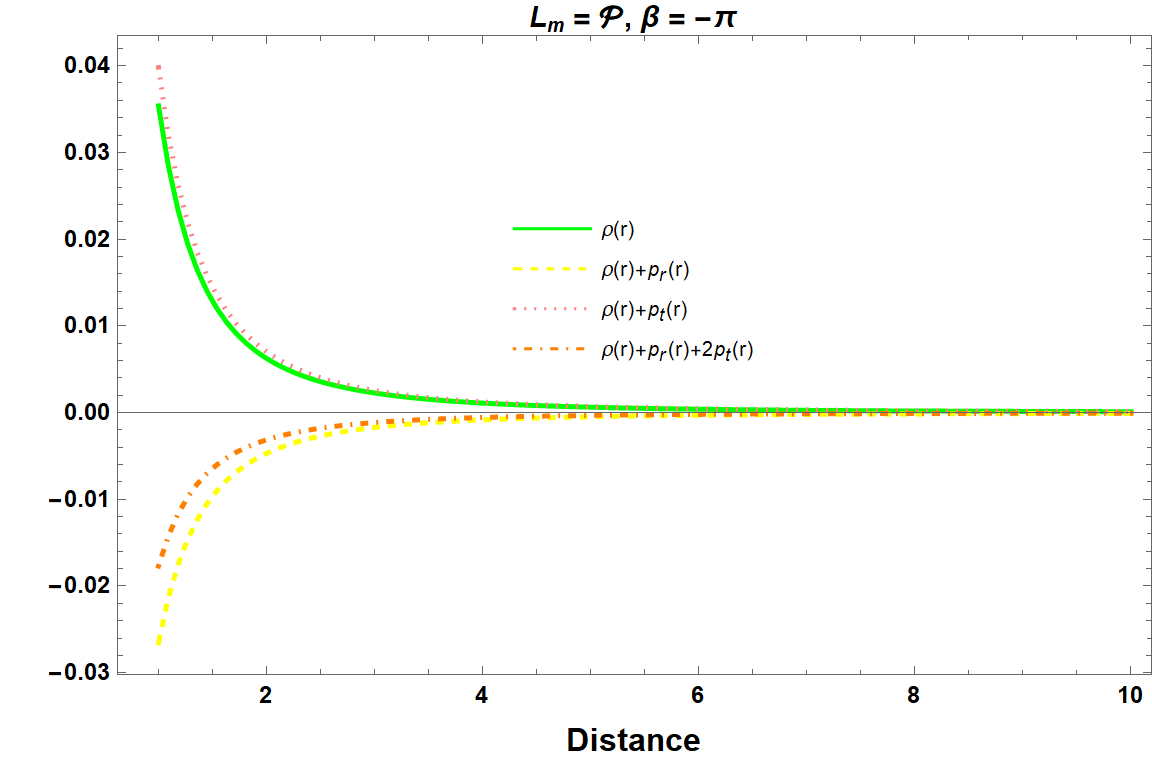}
    \includegraphics[width = 8.5 cm]{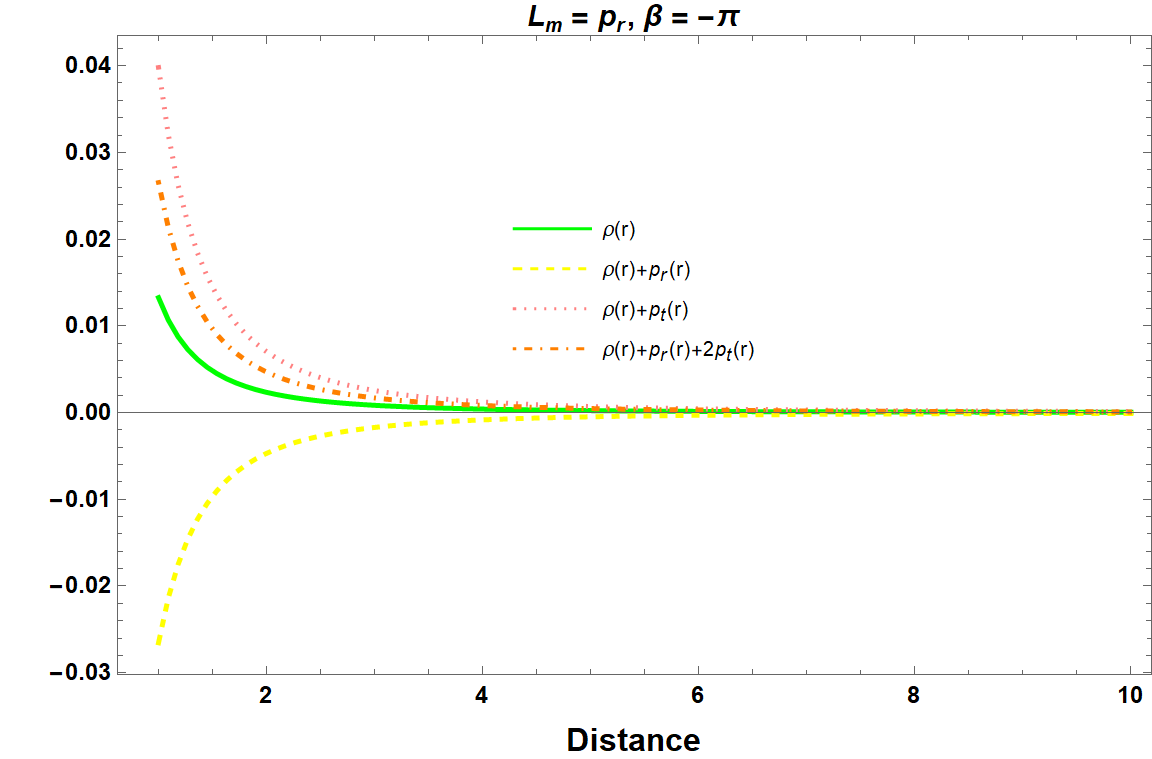}
\caption{The green-solid, yellow-dashed, pink-dotted and orange-dot-dashed 
curves denote the $\rho$ , $\rho+p_r$,  $\rho+p_t$ and $\rho+p_r+2p_t$, respectively for the specific case of $\Phi^{\prime}(r) = 0$ and $b(r)=  \sqrt{rr_0}$. In both
cases we consider $r_0=1$ and $\beta = -\pi$.
}\label{f1}
\end{figure}

Other choice for the Lagrangian density $\mathcal{L}_m = p_r$ and using the Eqs. (\ref{eq17}-\ref{eq19}),
the field equations are given by
\begin{eqnarray}
    \rho(r) &=& \frac{4\pi + 3 \beta}{8(2\pi + \beta)(4 \pi + \beta)} \frac{r_0}{r^3} \sqrt{\frac{r}{r_0}}, \\
    p_r(r) &=& -\frac{8\pi + 5 \beta}{8(2\pi + \beta)(4 \pi + \beta)} \frac{r_0}{r^3} \sqrt{\frac{r}{r_0}}, \\
    p_t(r) &=& \frac{\pi}{4(2 \pi + \beta)(4 \pi + \beta)}\frac{r_0}{ r^3} \sqrt{\frac{r}{r_0}}.
\end{eqnarray}
Searching for NEC along the radial and tangential direction which are given by
\begin{eqnarray}
\rho+p_r &=& -\frac{1 }{4 (4 \pi+\beta)}\frac{r_0}{r^3} \sqrt{\frac{r}{r_0}}, \\
     \rho+p_t  &=& \frac{3}{8 (4 \pi+\beta )}\frac{r_0}{r^3} \sqrt{\frac{r}{r_0}},
\end{eqnarray}
respectively. Using these expressions we plot $\rho$, $\rho+p_r$, $\rho+p_t$ and $\rho+p_r+2p_t$ in Fig. \ref{f1} (lower panel) for $r_0=1$ and $\beta = -\pi$. In this case, the energy density is positive throughout the spacetime also.  We see that $\rho+p_r <0$ and $\rho+p_t >0$. As in the previous case the situation 
is reversed when $\beta <-4 \pi$.  This means the violation of NEC for any values  of $\beta$.

On the other hand NEC at the throat is given by  
$(\rho+ p_r)|_{r_0} = -\frac{1 }{4r_0^2 (4 \pi+\beta )}$ and
$(\rho+p_r)|_{r_0} = \frac{3}{8 r_0^2 (4 \pi+\beta )}$, respectively. In this case also, the NEC is violated for the normal matter threading
the wormhole throat.

\vspace{0.5cm}


\subsection{Solutions with $\Phi(r)=\frac{r_0}{r}$ and $b(r)= r_0 \sqrt{\frac{r}{r_0}}$}

\begin{figure}[h]
    \centering
    \includegraphics[width = 8.5 cm]{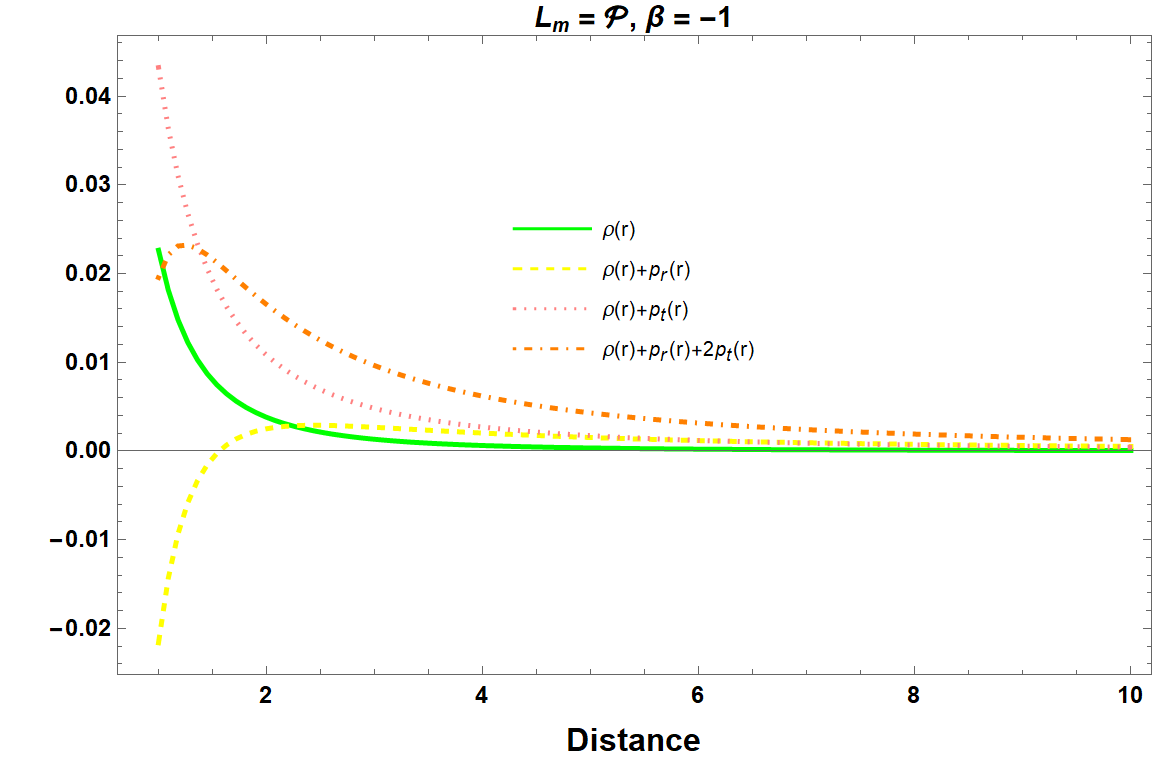}
    \includegraphics[width = 8.5 cm]{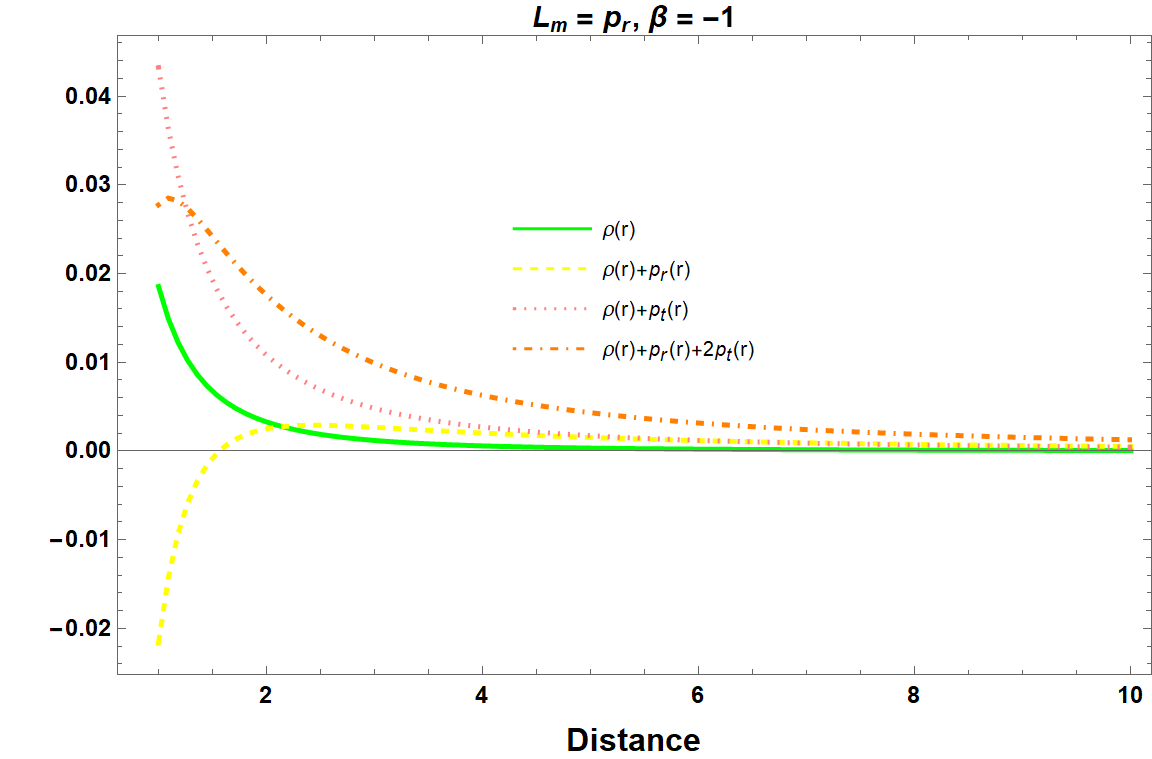}
    \caption{The green-solid, yellow-dashed, pink-dotted and orange-dot-dashed curves denote the $\rho$ , $\rho+p_r$,  $\rho+p_t$ and $\rho+p_r+2p_t$, respectively for the specific case of $\Phi^{\prime}(r) = 0$ and $b(r)=  \sqrt{rr_0}$. In both cases we consider $r_0=1$ and $\beta = -1$.}  \label{f3}  
\end{figure}

\subsubsection{For $\mathcal{L}_m = \mathcal{P}$}
Since the restricted choices for $\Phi(r)=\frac{r_0}{r}$ and $b(r)= r_0 \sqrt{\frac{r}{r_0}}$, and using the Eqs. (\ref{14}-\ref{16}), we obtain  

\begin{eqnarray}
        \rho(r) &=& \mathcal{K}_1 \left[\beta  \left(8 r^2+4 r_0^2 \mathcal{L}_1-r r_0\right)+24 \pi  r^2 \right], \label{eq33}\\
         p_r(r) &=& \mathcal{K}_1 \left[\beta  \left(-20 r^2-4 r_0^2 \mathcal{L}_1+r r_0 \left(1-48 \mathcal{L}_1\right)\right) \right. \nonumber\\
         && \left.
         -48 \pi  r \left(2 r_0 \mathcal{L}_1+r\right)\right], \label{eq34}\\
         p_t(r)&=& \mathcal{K}_1 \left[\beta  \left(10 r^2+20 r_0^2 \mathcal{L}_1+r r_0 \left(24 \mathcal{L}_1-5\right)\right) \right. \nonumber\\
         && \left.
         +12 \pi  \left(r^2+4 r_0^2 \mathcal{L}_1+r r_0 \left(4 \mathcal{L}_1-1\right)\right)\right]. \label{eq35}
\end{eqnarray}
$\mathcal{K}_1 = \frac{1}{48 (\beta +2 \pi ) (\beta +4 \pi ) r^4 \sqrt{\frac{r}{r_0}}}$ and $\mathcal{L}_1 = \left(\sqrt{\frac{r}r_0}-1\right)$. Now, with the help of the above Eqs. (\ref{eq33})-(\ref{eq35}),  we plot $\rho$, $\rho+p_r$, $\rho+p_t$ and $\rho+p_r+2p_t$
in Fig. \ref{f3} (upper panel). The energy conditions are realized by choosing a particular value of $r_0=1$ and $\beta = -1$. Fig. \ref{f3} (upper panel) shows that energy density is positive
throughout the spacetime, while $\rho+p_r <0$. It is worth to mention that this model also violate the NEC.
Here, we have explicitly written the expression for $\rho+p_r$ and $\rho+p_t$ at the throat which are given by
\begin{eqnarray}
    (\rho+p_r)|_{r_0} &=& -\frac{1}{4r_0^2 (4 \pi+ \beta ) } , \label{eq36} \\
     (\rho+p_t)|_{r_0}  &=&  \frac{1}{4r_0^2 (4 \pi+ \beta ) }, \label{eq37} 
\end{eqnarray}
respectively. In this context, we see that NEC is violated for
the normal matter threading the wormhole throat. More specifically, for the choice of $\beta \lessgtr 4\pi$ the $\rho(r_0)+p_r(r_0) \lessgtr 0$. This means that the situation does not change depending on the parameter $\beta$, see Eqs. (\ref{eq36}) and (\ref{eq37}).

\subsubsection{For $\mathcal{L}_m = p_r$}
Adopting the Lagrangian density $\mathcal{L}_m = p_r$ and using the set of equations (\ref{eq17})-(\ref{eq19}), the energy density, radial pressure and the transverse pressure are given by the following expressions
\begin{widetext}
   \begin{eqnarray}
        \rho(r) &=&  \mathcal{K}_2 \left[\beta  \left(r^2 \left(6 \mathcal{L}_2+8\right)-4 r_0^2 \mathcal{L}_2+r r_0 \left(4-9 \mathcal{L}_2\right)\right)+8 \pi  r^2 \mathcal{L}_2\right], \label{eq38}\\
         p_r(r) &=& \mathcal{K}_2 \left[\beta  \left(-2 r^2 \left(5 \mathcal{L}_2+12\right)+4 r_0^2 \mathcal{L}_2+r r_0 \left(25 \mathcal{L}_2-4\right)\right)-16 \pi  r \left(r \left(\mathcal{L}_2+2\right)-2 r_0 \mathcal{L}_2\right)\right], \label{eq39}\\
         p_t(r)&=& \mathcal{K}_2 \left[ 4 \pi  \left(r^2 \left(\mathcal{L}_2+4\right)-4 r_0^2 \mathcal{L}_2+r r_0 \left(4-5 \mathcal{L}_2\right)\right)-\beta  r_0 \left(r \left(\mathcal{L}_2-4\right)+4 r_0 \mathcal{L}_2\right)\right]. \label{eq40}
\end{eqnarray}
\end{widetext}
where $\mathcal{K}_2 = \frac{r_0}{16 (\beta +2 \pi ) (\beta +4 \pi ) r^5}$ and $\mathcal{L}_2 = \sqrt{\frac{r}{r_0}}$. Setting $r_0=1$ and $\beta = -1$, we examine the NEC, WEC and SEC in Fig. \ref{f3} (lower panel). Now, in order to check the NEC, we see that the energy density is positive throughout the spacetime, while $\rho+p_r <0$. From our analysis it is clear that the NEC (and also WEC)
is violated as depicted in Fig. \ref{f3} (lower panel). To check the NEC of matter threading the wormhole at the throat given by 
\begin{eqnarray}
    (\rho+p_r)|_{r_0} &=& -\frac{1}{4 {r_0}^2(4 \pi+ \beta )} , \label{eq41} \\
     (\rho+p_t)|_{r_0}  &=&  \frac{1}{4 r_0^2(4 \pi + \beta) }, \label{eq42} 
\end{eqnarray}   
Interestingly, the Eqs. (\ref{eq41})-(\ref{eq42}) are same as of Eqs. (\ref{eq36})-(\ref{eq37}). Thus, the NEC and consequently the WEC is violated at the throat.


\subsection{Isotropic case $p_r = p_t = p$}

Consider an isotropic pressure, $p_r = p_t = p$, we have three equations with four unknowns, namely, $\rho(r)$, $p(r)$, $\Phi(r)$ and $b(r)$. Here, we analyze the wormhole with a constant redshift function, $\Phi^{\prime}(r)= 0$. Considering this specific choice, one obtains the same set of field equations for
(\ref{14})-(\ref{16}) and (\ref{eq17})-(\ref{eq19}), which are
\begin{eqnarray}
     \frac{b^{\prime}}{r^{2}} &=& 8\pi\rho + \beta\left( 3\rho - p  \right) , \label{eq43} \\
     -\frac{b}{r^{3}} &=& 8\pi p + \beta\left( -\rho + 3p  \right), \label{eq44} \\
      -\frac{b^{\prime}r-b}{2r^3}  &=& 8\pi p +   \beta\left( -\rho + 3p \right). \label{eq45}
\end{eqnarray}
In this case, Eqs. (\ref{eq44}) and  (\ref{eq45}) leads the following shape function $b(r)=Cr^3$. Satisfying the value $b(r_0) = r_0 $, the shape function yields $b(r)= \frac{r^3}{r_0^2}$, and  
the spacetime metric takes the final form
\begin{equation}\label{eq13}
ds^2=- dt^2+\frac{dr^2}{1- \frac{r^3}{r_0^2}}+r^2(d\theta^2+\sin^2\theta d\phi^2).
\end{equation}
As we know a necessary condition for wormhole geometry is the shape function $b(r)$ must obey the condition i.e., $b'(r_0) < 1$. In our case we found that $b'(r_0) =3 \nless 1$. This implies that wormhole solution does not satisfy the flaring out condition at the throat 
for perfect fluid in $f(R,T)$ gravity.  Moreover, the asymptotically flatness condition is also violating. Interestingly, we obtain the same results for  reported in \cite{Banerjee:2020uyi}.

\section{Concluding remarks}\label{sec6}

In this work, we have explored wormhole geometries in $f(R,T)$ gravity.  An interesting feature of this theory is the curvature-matter coupling  i.e., the gravitational Lagrangian of $f(R,T)$ theory is an arbitrary function of the Ricci scalar and the trace of the stress-energy tensor. Here, we turn our attention to the construction of wormhole solution for two different choices of the matter Lagrangian density 
($\mathcal{L}_m= \mathcal{P}$ versus $\mathcal{L}_m= p_r$) within the framework of $f(R,T)$ gravity. More specifically, we obtain exact wormhole solutions by specifying 
the choice for a redshift function and a shape function. The main interest in our solution is to examine whether the matter field obeys NEC everywhere including the throat or not. 

Consider a linear form function $f(R,T)= R+ 2\beta T$,  we obtain two different sets of gravitational field equations
(\ref{14}-\ref{16}) and (\ref{eq17}-\ref{eq19}), respectively. 
Next we consider the specific choice of redshift and shape function to solve the field equations. Considering the 
field equations with a constant $\Phi(r)$ and $b(r)= \sqrt{r_0 r}$, we theoretically construct wormhole solution. In this context, we see wormhole solutions violating NEC throughout the spacetime i.e., $\rho+p_r <0$ and $\rho+p_t >0$ for $r_0=1$ and $\beta = -\pi$ in both cases, as can be readily verified from Fig. \ref{f1}. Secondly, we obtain further wormhole geometries by considering $\Phi(r)=\frac{r_0}{r}$ and $b(r)= \sqrt{r_0 r}$. It is interesting to see that the NEC (and consequently the WEC) is also violated, i.e., $\rho+p_r <0$ by setting $r_0=1$ and $\beta = -1$. One can verify this situation from  Fig. \ref{f3}. Finally, we study wormholes sustained by  matter sources with isotropic pressure. In this case we show that wormhole solutions could not sustain for a zero-tidal-force. This result is consistent with the solution obtained in \cite{Banerjee:2020uyi}.
 

\section*{Acknowledgments}
T.~Tangphati is supported by School of Science, Walailak University, Thailand. A. Pradhan thanks to IUCCA, Pune, India for providing facilities under associateship programmes.


\begin{thebibliography}{0}%
\makeatletter
\providecommand \@ifxundefined [1]{%
 \@ifx{#1\undefined}
}%
\providecommand \@ifnum [1]{%
 \ifnum #1\expandafter \@firstoftwo
 \else \expandafter \@secondoftwo
 \fi
}%
\providecommand \@ifx [1]{%
 \ifx #1\expandafter \@firstoftwo
 \else \expandafter \@secondoftwo
 \fi
}%
\providecommand \natexlab [1]{#1}%
\providecommand \enquote  [1]{``#1''}%
\providecommand \bibnamefont  [1]{#1}%
\providecommand \bibfnamefont [1]{#1}%
\providecommand \citenamefont [1]{#1}%
\providecommand \href@noop [0]{\@secondoftwo}%
\providecommand \href [0]{\begingroup \@sanitize@url \@href}%
\providecommand \@href[1]{\@@startlink{#1}\@@href}%
\providecommand \@@href[1]{\endgroup#1\@@endlink}%
\providecommand \@sanitize@url [0]{\catcode `\\12\catcode `\$12\catcode
  `\&12\catcode `\#12\catcode `\^12\catcode `\_12\catcode `\%12\relax}%
\providecommand \@@startlink[1]{}%
\providecommand \@@endlink[0]{}%
\providecommand \url  [0]{\begingroup\@sanitize@url \@url }%
\providecommand \@url [1]{\endgroup\@href {#1}{\urlprefix }}%
\providecommand \urlprefix  [0]{URL }%
\providecommand \Eprint [0]{\href }%
\providecommand \doibase [0]{https://doi.org/}%
\providecommand \selectlanguage [0]{\@gobble}%
\providecommand \bibinfo  [0]{\@secondoftwo}%
\providecommand \bibfield  [0]{\@secondoftwo}%
\providecommand \translation [1]{[#1]}%
\providecommand \BibitemOpen [0]{}%
\providecommand \bibitemStop [0]{}%
\providecommand \bibitemNoStop [0]{.\EOS\space}%
\providecommand \EOS [0]{\spacefactor3000\relax}%
\providecommand \BibitemShut  [1]{\csname bibitem#1\endcsname}%
\let\auto@bib@innerbib\@empty
\end{thebibliography}%


\begin{thebibliography}{}


\bibitem{Einstein:1953tkd}
A.Einstein and N. Rosen. 
Phys. Rev., \textbf{48}, 73, (1953).

\bibitem{Fuller1957} 
J. A. Wheeler,  Annals Phys. \  {\bf 2},  604 (1957).

\bibitem{Morris:1988cz} 
M.~S.~Morris and K.~S.~Thorne,
Am.\ J.\ Phys.\  {\bf 56}, (1988) 395.

\bibitem{Morris1988}  
M. S. Morris, K. S. Thorne and U. Yurtsever, 
Phys. Rev. Lett. \textbf{ 61}, 1446 (1988).

\bibitem{Visser:1995} 
M. Visser, Lorentzian Wormholes: From Einstein to
Hawking, (American Institute of Physics, New York,
1995).

\bibitem{Lobo:2007zb}
F.~S.~N.~Lobo,
arXiv:0710.4474 [gr-qc].

\bibitem{Visser:2003yf}
M.~Visser, S.~Kar and N.~Dadhich,
Phys. Rev. Lett. \textbf{90}, 201102 (2003).

\bibitem{Kar:2004hc}
S.~Kar, N.~Dadhich and M.~Visser,
Pramana \textbf{63}, 859 (2004).

\bibitem{Nandi:2004ku}
K.~K.~Nandi, Y.~Z.~Zhang and K.~B.~Vijaya Kumar,
Phys. Rev. D \textbf{70}, 127503 (2004).

\bibitem{Poisson:1995sv}
E.~Poisson and M.~Visser,
Phys. Rev. D \textbf{52}, 7318 (1995).

\bibitem{Visser:1989kh}
 M.~Visser,
Phys.\ Rev.\ D  {\bf 39} (1989) 3182.

\bibitem{Visser:1989kg}
 M.~Visser,
 Nucl.\ Phys.\ B {\bf 328} (1989) 203.
		
\bibitem{Lobo:2003xd}
F.~S.~N.~Lobo and P.~Crawford,
Class. Quant. Grav. \textbf{21}, 391 (2004).

\bibitem{Dias:2010uh}
G.~A.~S.~Dias and J.~P.~S.~Lemos,
Phys. Rev. D \textbf{82}, 084023 (2010).

\bibitem{Lobo:2005us}
F.~S.~N.~Lobo,
Phys. Rev. D \textbf{71}, 084011 (2005).

\bibitem{Sushkov:2005kj}
S.~V.~Sushkov,
Phys. Rev. D \textbf{71}, 043520 (2005).


\bibitem{Carvente:2019gkd}
B.~Carvente, V.~Jaramillo, J.~C.~Degollado, D.~N\'u\~nez and O.~Sarbach,
Class. Quant. Grav. \textbf{36},  235005 (2019).

\bibitem{Sharif:2014opa}
M.~Sharif and A.~Jawad,
Eur. Phys. J. Plus \textbf{129}, 15 (2014).

\bibitem{Jamil:2010ziq}
M.~Jamil, P.~K.~F.~Kuhfittig, F.~Rahaman and S.~A.~Rakib,
Eur. Phys. J. C \textbf{67}, 513 (2010).


\bibitem{Zaslavskii:2005fs}
O.~B.~Zaslavskii,
Phys. Rev. D \textbf{72}, 061303 (2005).

\bibitem{Bronnikov:2006pt}
K.~A.~Bronnikov and A.~A.~Starobinsky,
JETP Lett. \textbf{85}, 1 (2007).

\bibitem{Gonzalez:2009cy}
J.~A.~Gonzalez, F.~S.~Guzman, N.~Montelongo-Garcia and T.~Zannias,
Phys. Rev. D \textbf{79}, 064027 (2009).

\bibitem{Cataldo:2008ku}
M.~Cataldo, S.~del Campo, P.~Minning and P.~Salgado,
Phys. Rev. D \textbf{79}, 024005 (2009).

\bibitem{Zangeneh:2014noa}
M.~K.~Zangeneh, F.~S.~N.~Lobo and N.~Riazi,
Phys. Rev. D \textbf{90}, 024072 (2014).

\bibitem{Mazharimousavi:2016npo}
S.~H.~Mazharimousavi and M.~Halilsoy,
Mod. Phys. Lett. A \textbf{31},  1650192 (2016).

\bibitem{Bronnikov:2002rn}
K.~A.~Bronnikov and S.~W.~Kim,
Phys. Rev. D \textbf{67}, 064027 (2003)

\bibitem{Mehdizadeh:2015jra}
M.~R.~Mehdizadeh, M.~Kord Zangeneh and F.~S.~N.~Lobo,
Phys. Rev. D \textbf{91},  084004 (2015).



\bibitem{Pavlovic:2014gba}
P.~Pavlovic and M.~Sossich,
Eur. Phys. J. C \textbf{75}, 117 (2015).

\bibitem{Lobo:2009ip}
F.~S.~N.~Lobo and M.~A.~Oliveira,
Phys. Rev. D \textbf{80}, 104012 (2009).

\bibitem{Jusufi:2020yus}
K.~Jusufi, A.~Banerjee and S.~G.~Ghosh,
Eur. Phys. J. C \textbf{80}, 698 (2020). 

\bibitem{Godani:2022jwz}
N.~Godani, D.~V.~Singh and G.~C.~Samanta,
Phys. Dark Univ. \textbf{35}, 100952 (2022). 

\bibitem{KordZangeneh:2015dks}
M.~Kord Zangeneh, F.~S.~N.~Lobo and M.~H.~Dehghani,
Phys. Rev. D \textbf{92},  124049 (2015).

\bibitem{Mehdizadeh:2016nna}
M.~R.~Mehdizadeh and F.~S.~N.~Lobo,
Phys. Rev. D \textbf{93},  124014 (2016).

\bibitem{Bakopoulos:2021liw}
A.~Bakopoulos, C.~Charmousis and P.~Kanti,
JCAP \textbf{05},  022 (2022).

\bibitem{Rosa:2021yym}
J.~L.~Rosa,
Phys. Rev. D \textbf{104},  064002 (2021).

\bibitem{KordZangeneh:2020ixt}
M.~Kord Zangeneh and F.~S.~N.~Lobo,
Eur. Phys. J. C \textbf{81},  285 (2021).

\bibitem{Banerjee:2021mqk}
A.~Banerjee, A.~Pradhan, T.~Tangphati and F.~Rahaman,
Eur. Phys. J. C \textbf{81}, 1031 (2021).

\bibitem{Parsaei:2022wnu}
F.~Parsaei, S.~Rastgoo and P.~K.~Sahoo,
Eur. Phys. J. Plus \textbf{137}, 1083 (2022).

\bibitem{Hassan:2022hcb}
Z.~Hassan, S.~Ghosh, P.~K.~Sahoo and K.~Bamba,
Eur. Phys. J. C \textbf{82}, 1116 (2022).

\bibitem{DeFalco:2021klh}
V.~De Falco, E.~Battista, S.~Capozziello and M.~De Laurentis,
Phys. Rev. D \textbf{103},  044007 (2021).

\bibitem{DeFalco:2021ksd}
V.~De Falco, E.~Battista, S.~Capozziello and M.~De Laurentis,
Eur. Phys. J. C \textbf{81},  157 (2021).

\bibitem{Harko:2011kv}
T.~Harko, F.~S.~N.~Lobo, S.~Nojiri and S.~D.~Odintsov,
Phys. Rev. D \textbf{84}, 024020 (2011).

\bibitem{Houndjo:2011fb} 
  M.~J.~S.~Houndjo and O.~F.~Piattella,
   Int.\ J.\ Mod.\ Phys.\ D {\bf 21} (2012) 1250024.
  
  \bibitem{jamil2011re}
  M.~Jamil, D.~Momeni, M.~Raza and R.~Myrzakulov,
   Eur.\ Phys.\ J.\ C {\bf 72} (2012) 1999.
  
  \bibitem{Jamil:2012pf}
  M.~Jamil, D.~Momeni and R.~Myrzakulov,
   Chin.\ Phys.\ Lett.   {\bf 29} (2012) 109801.
  
  \bibitem{Baffou:2013dpa}
  E.~H.~Baffou, A.~V.~Kpadonou, M.~E.~Rodrigues, M.~J.~S.~Houndjo and J.~Tossa,
   Astrophys.\ Space Sci.\   {\bf 356} (2015) 173.
  
  \bibitem{Singh:2013bpa}
  C.~P.~Singh and V.~Singh,
   Gen.\ Rel.\ Grav.\  {\bf 46} (2014) 1696.
  
  \bibitem{Sharif:2014ioa} 
  M.~Sharif and M.~Zubair,
   Gen.\ Rel.\ Grav.\  {\bf 46} (2014) 1723.
  
  \bibitem{Baffou:2017pao}
  E.~H.~Baffou, M.~J.~S.~Houndjo, M.~Hamani-Daouda and F.~G.~Alvarenga,
   Eur.\ Phys.\ J.\ C {\bf 77} (2017) 708.
  
  \bibitem{Mishra:2017sdq}
  B.~Mishra, S.~Tarai and S.~K.~Tripathy,
   Mod.\ Phys.\ Lett.\ A {\bf 33}, 1850170 (2018). 

\bibitem{Noureen:2014xua}
I.~Noureen and M.~Zubair,
Astrophys. Space Sci. \textbf{356},  103 (2015).

\bibitem{Moraes2016}
P.~H.~R.~S.~Moraes, J.~D.~V.~Arba\~nil and M.~Malheiro,
JCAP \textbf{06}, 005 (2016).

\bibitem{Das:2016mxq}
A.~Das, F.~Rahaman, B.~K.~Guha and S.~Ray,
Eur. Phys. J. C \textbf{76}, 654 (2016).

\bibitem{Deb:2017rhd}
D.~Deb, F.~Rahaman, S.~Ray and B.~K.~Guha,
JCAP \textbf{03}, 044 (2018)

\bibitem{Biswas:2018inc}
S.~Biswas et al.,
Annals Phys. \textbf{401}, 1 (2019).

\bibitem{Lobato2020}
R. Lobato et al.,
JCAP \textbf{12}, 039 (2020).  

\bibitem{Pretel2021}
J. M. Z. Pretel, S. E. Jor\'as, R. R. R. Reis and J. D. V. Arba\~nil,
JCAP \textbf{08}, 055 (2021).  

\bibitem{Pappas:2022gtt}
T.~D.~Pappas, C.~Posada and Z.~Stuchl\'\i{}k,
Phys. Rev. D \textbf{106}, 12 (2022).

\bibitem{Ordines:2019sjq}
T.~M.~Ordines and E.~D.~Carlson,
Phys. Rev. D \textbf{99}, 104052 (2019).

\bibitem{Moraes:2017mir}
P.~H.~R.~S.~Moraes and P.~K.~Sahoo,
Phys. Rev. D \textbf{96},  044038 (2017).

\bibitem{Elizalde:2018frj}
E.~Elizalde and M.~Khurshudyan,
Phys. Rev. D \textbf{98},  123525 (2018).

\bibitem{Moraes:2019pao}
P.~H.~R.~S.~Moraes and P.~K.~Sahoo,
Eur. Phys. J. C \textbf{79}, 677 (2019).

\bibitem{Zubair:2019uul}
M.~Zubair, R.~Saleem, Y.~Ahmad and G.~Abbas,
Int. J. Geom. Meth. Mod. Phys. \textbf{16},  1950046 (2019).

\bibitem{Rosa:2022osy}
J.~L.~Rosa and P.~M.~Kull,
Eur. Phys. J. C \textbf{82}, 1154 (2022).

\bibitem{Banerjee:2020uyi}
A.~Banerjee, M.~K.~Jasim and S.~G.~Ghosh,
Annals Phys. \textbf{433}, 168575 (2021).

\bibitem{Moraes:2017rrv} 
P.~H.~R.~S.~Moraes, W.~de Paula and R.~A.~C.~Correa,
Int.\ J.\ Mod.\ Phys.\ D {\bf 28},  1950098 (2019).

\bibitem{Banerjee:2019wjj}
A.~Banerjee, K.~N.~Singh, M.~K.~Jasim and F.~Rahaman,
Annals Phys. \textbf{422}, 168295 (2020).













\bibitem{Sotiriou:2008rp}
T.~P.~Sotiriou and V.~Faraoni,
Rev. Mod. Phys. \textbf{82}, 451 (2010).

\bibitem{DeFelice:2010aj}
A.~De Felice and S.~Tsujikawa,
Living Rev. Rel. \textbf{13}, 3 (2010).

\bibitem{Nojiri:2010wj}
S.~Nojiri and S.~D.~Odintsov,
Phys. Rept. \textbf{505}, 59 (2011).

\bibitem{Nojiri:2017ncd}
S.~Nojiri, S.~D.~Odintsov and V.~K.~Oikonomou,
Phys. Rept. \textbf{692}, 1 (2017).

\bibitem{BarrientosO:2014mys}
J.~Barrientos O. and G.~F.~Rubilar,
Phys. Rev. D \textbf{90},  028501 (2014).

\bibitem{Harko:2010zi}
T.~Harko,
Phys. Rev. D \textbf{81}, 044021 (2010).

\bibitem{Faraoni:2009rk}
V.~Faraoni,
Phys. Rev. D \textbf{80}, 124040 (2009).

\bibitem{Bertolami:2008ab}
O.~Bertolami, F.~S.~N.~Lobo and J.~Paramos,
Phys. Rev. D \textbf{78}, 064036 (2008).

\bibitem{Deb:2018sgt}
D.~Deb, S.~V.~Ketov, S.~K.~Maurya, M.~Khlopov, P.~H.~R.~S.~Moraes and S.~Ray,
Mon. Not. Roy. Astron. Soc. \textbf{485},5652 (2019).

\bibitem{Maurya:2019sfm}
S.~K.~Maurya, A.~Errehymy, D.~Deb, F.~Tello-Ortiz and M.~Daoud,
Phys. Rev. D \textbf{100}, 044014 (2019).

\bibitem{Biswas:2020gzd}
S.~Biswas, D.~Shee, B.~K.~Guha and S.~Ray,
Eur. Phys. J. C \textbf{80},  175 (2020).

\bibitem{Bouhmadi-Lopez:2014gza}
M.~Bouhmadi-L\'opez, F.~S.~N.~Lobo and P.~Mart\'\i{}n-Moruno,
JCAP \textbf{11}, 007 (2014).





\end{thebibliography}
\end{document}